\documentstyle[12pt,aasms4] {article}
\def\pa{\partial}

\def\t{{\rm t}}

\newcommand{\beq}{\begin{equation}}
\newcommand{\eeq}{\end{equation}}
\newcommand{\beqn}{\begin{eqnarray}}
\newcommand{\eeqn}{\end{eqnarray}} 
\newcommand{\non}{\nonumber}

\def\t0{\theta_{\circ}}

\def\be{\begin{equation}}
\def\en{\end{equation}}

\def\gapp{\ \lower 3pt\hbox{${\buildrel > \over \sim}$}\ }
\def\lapp{\ \lower 3pt\hbox{${\buildrel < \over \sim}$}\ }
\newcommand{\no}{\noindent}
\begin{document}
\title{On the Chermnykh-Like Problems:\\
       I. The Mass Parameter $\mu=0.5$}

\author{Ing-Guey Jiang$^{1}$ and Li-Chin Yeh$^{2}$ }
        

\affil{{$^{1}$ Institute of Astronomy,}
{ National Central University, Chung-Li, Taiwan} \\
{$^{2}$ Department of Applied Mathematics,}
{ National Hsinchu University of Education, Hsin-Chu, Taiwan} \\}

\date{Received ?; accepted ?}

\begin{abstract} {

Following Papadakis (2005)'s numerical exploration of the Chermnykh's problem,
we here study
a Chermnykh-like problem motivated by the astrophysical applications. 
We find that both the equilibrium
points and solution curves become quite different from the ones of 
the classical planar
restricted three-body problem.
In addition to the usual Lagrangian points, there are new 
equilibrium points in our system.
We also calculate the Lyapunov Exponents for some 
example orbits.
We conclude that it seems there are more chaotic 
orbits for the system  
when there is a belt to interact with.
 }
\end{abstract}

%

\section{Introduction}

Chermnykh's problem concerns the motion of a test particle 
in the orbital plane of a dumb-bell. This dumb-bell rotates with constant
angular velocity $n$ around its mass center.
The equations of motion of this test particle are (Chermnykh 1987):
\beqn
&&\frac{dx}{dt}=u \non \\
&&\frac{dy}{dt}=v \non \\
&&\frac{du}{dt}=2nv-\frac{\pa U^{\ast}}{\pa x}\non  \\
&& \frac{dv}{dt}=-2nu-\frac{\pa U^{\ast}}{\pa y},\non \\
\label{eq:cher1} 
\eeqn
where the potential $U^{\ast}$ is 
\beq
U^{\ast}=-\frac{n^2}{2}(x^2+y^2)-\frac{\mu_1}{r_1}-\frac{\mu_2}{r_2},
\label{eq:u_ast}
\eeq
$r_1=\sqrt{(x+\mu_2)^2+y^2}$, $r_2=\sqrt{(x-\mu_1)^2+y^2}$
and $\mu_1+\mu_2=1$. Usually, the mass parameter $\mu \equiv \mu_2$ and
the parameter $n \in [0, \infty)$.

This interesting problem was first analysed in Chermnykh (1987) on 
the Lyapunov stability of the triangular solutions. 
Gozdziewski and Maciejewski (1998) studied the Lyapunov stability of 
the Lagrangian points in the entire range of angular velocity $n$
and mass parameter $\mu$. Recently, Papadakis (2005) investigated
the equilibrium points and the zero-velocity curves of this problem. 
In particular, some families of periodic orbits are studied in great detail.
The stability zones are also discussed in that paper.

In general, this important 
problem has many applications in celestial mechanics (see 
Gozdziewski and Maciejewski 1999) 
and chemistry (see Strand and Reinhardt 1979).                .
When the angular velocity $n=1$, it is the classical planar 
restricted three-body problem. When $n=0$, we get the Euler's problem
of two fixed gravitational centers. 

On the other hand,
because (i) there are many discovered extra-solar planetary systems
(Laughlin \& Adams 1999, Rivera \& Lissauer 2000, Jiang \& Ip 2001,
Ji et al. 2002), in 
which about five of them are in binary stars and, 
(ii) the effect of belts might be
important according to some of our previous studies (Yeh \& Jiang 2001,
Jiang \& Yeh 2003, 2004), we here construct
a Chermnykh-like problem motivated by these astrophysical applications.
In principle, we consider the angular velocity $n>1$
and its value is 
determined by the belt's gravitational potential, which is considered as part
of the total potential to govern the test particle's motion. 
 
Therefore, we construct our basic model in Section 2. In Section 3,
we study the existence of new equilibrium points. We discuss the behavior
of the orbits with initial conditions near the new equilibrium points
in Section 4 and calculate their values of 
Lyapunov Exponent in Section 5. Section 6 concludes the paper.

\section{The Model}

We consider the motion of a test particle influenced by 
the gravitational force from the central binary and the circumbinary disc.

For convenience, we set the gravitational constant $G=1$. We  
assume that two masses of the central binary 
are $\mu_1$ and $\mu_2$ and choose
the unit of mass to make $\mu_1+\mu_2=1$. 
The separation between two stars in the central binary is set to be unity.
The time unit is therefore determined from the above choice.  

Under the influence from the belt, when $\mu_1=\mu_2=0.5$,
both components of the central binary
move on circular orbits at $r=0.5$ 
with mean motion, i.e. angular velocity $n={\sqrt {1-2 f_b(0.5)}}$.
The gravitational force $f_b$ from the belt is
\beq
f_b(r)=-\frac{\pa V}{\pa r}= 
-2\int^{r_o}_{r_i}\frac{\rho(r')r'}{r}\left[\frac{E(\xi)}{r-r'}+
\frac{F(\xi)}{r+r'}\right] dr',\label{eq:fb}
\eeq
where 
\beq
\xi=\frac{2\sqrt{rr'}}{r+r'},\label{eq:xi}
\eeq
$F(\xi)$ and $E(\xi)$ are elliptic 
integral of the first kind and the second kind and 
\beqn
F(\xi)&=&\int^{\pi/2}_0\frac{1}{\sqrt{1-\xi^2\sin^2\phi'}}d\phi',\non \\
E(\xi)&=&\int^{\pi/2}_0 {\sqrt{1-\xi^2\sin^2\phi'}}d\phi'.
\eeqn
(Please see Appendix A for the derivation of Eq.(\ref{eq:fb}).)
Hence, the $x$ and $y$ components of gravitational force are 
\beqn
&& -\frac{\pa V}{\pa x}= f_b\frac{x}{r},  \non \\
&&-\frac{\pa V}{\pa y}= f_b\frac{y}{r}, \non \\
\label{eq:pav}
\eeqn
where $f_b$ is in Eq.(\ref{eq:fb}) and $V$ is the potential from the belt.  

Because both components of central binary are  
doing circular motions, the equation of motion
can be written on the frame rotating with the central binary. 
The equations of motion of this problem are 
\beqn
&&\frac{dx}{dt}=u \non \\
&&\frac{dy}{dt}=v \non \\
&&\frac{du}{dt}=2nv-\frac{\pa U^{\ast}}{\pa x}-\frac{\pa V}{\pa x}\non  \\
&& \frac{dv}{dt}=-2nu-\frac{\pa U^{\ast}}{\pa y}-\frac{\pa V}{\pa y},\non \\
\label{eq:3body1} 
\eeqn
where the potential $U^{\ast}$ is given by Eq.(2).
 $r_1=\sqrt{(x+\mu_2)^2+y^2}$ and $r_2=\sqrt{(x-\mu_1)^2+y^2}$.
The Jacobi integral for this system is therefore
\beq
C_J = -u^2 -v^2 -2U^{\ast} -2V.
\eeq

The density profile of the belt is 
\beq
\rho(r)=\left\{\begin{array}{ll}
0  & {\rm  when}\,\,  r < r_{i} \,\,{\rm or}\,\,  r > r_{o},\\
 \frac{c}{r^p}\left\{\cos\left[ \frac{\pi}{2} \frac{(r-r_a)}{(r_{i} - r_a)} 
\right]\right\}^2 & 
{\rm  when} \,\,   r_{i} < r < r_a,  \\
 \frac{c}{r^p} & {\rm when} \,\,   r_a < r < r_b,  \\
\frac{c}{r^p}\left\{\cos\left[ \frac{\pi}{2} \frac{(r-r_b)}{(r_{o}-r_b)} 
\right]\right\}^2 & 
{\rm  when} \,\,   r_b < r < r_{o},  \\
\end{array}\right.
\eeq
where 
$r=\sqrt{x^2+y^2}$, $c$ is a constant 
completely determined by the total mass of the belt and $p$ is a natural 
number. In this paper, we set $p=2$, $r_a=r_i+0.1$, $r_b=r_i+0.9$, 
and $r_o=r_i+1.0$ for all numerical results 
(Lizano \& Shu 1989).
It is a power-law profile with smooth
edges.
Hence, the total mass of the belt is 
\beqn
M_{b}&=&\int^{2\pi}_{0}\int^{r_o}_{r_i}\rho(r')r'dr'd\phi \non\\
&=&2\pi c
\left\{\int^{r_a}_{r_i}\frac{1}{r'}\left\{\cos\left[ \frac{\pi}{2} 
\frac{(r-r_a)}{(r_{i} - r_a)} \right]\right\}^2 dr'+
\ln (r_b/r_a)   \right. \non \\
&+&  \left. \int^{r_o}_{r_b}\frac{1}{r'}\left\{\cos\left[ \frac{\pi}{2} 
\frac{(r-r_b)}{(r_{o}-r_b)} \right]\right\}^2 dr'\right\}. 
\eeqn
We use $r_i$ and $M_b$ as the controlling parameters of the belt profile.
Using Eq.(8), we determine the zero-velocity curves of our system
when $r_i=0.7$ and $M_b=0.3$,
as shown in Figure 1. Because there are elliptic integrals in the potential
$V$, this has to be done numerically and the points are not uniformly 
distributed on the curves.
For the results 
in Section 4 and Section 5, we always set $r_i=0.7$ but $M_b=0.3$ or 0.
Following Tancredi et al. (2001), 
we will integrate the orbits up to 1000 binary period.

\section{The Equilibrium Points}

We know that there are five equilibrium points (Lagrange points) 
for the classical restricted three-body problem.  Thus, our system 
shall have five
equilibrium points when $M_b=0$. It would be interesting
to investigate the number of equilibrium points when $M_b > 0$. When there 
are more than five equilibrium points, we claim that the new equilibrium 
points exist. 

The equilibrium points $(x_e,y_e)$ of System (\ref{eq:3body1}) satisfy the 
following equations:
\beqn 
f(x,y)&\equiv& n^2 x-\frac{\mu_1(x+\mu_2)}{r_1^3}-\frac{\mu_2(x-\mu_1)}
{r_2^3} +\frac{x}{r}f_b(r)=0,\label{eq:gg1}  \\
g(x,y)&\equiv& n^2 y-\frac{y\mu_1}{r_1^3}-\frac{y\mu_2}{r_2^3}
+\frac{y}{r}f_b(r)=0,\label{eq:gg2}
\eeqn

Because we consider $\mu_1=\mu_2=0.5$ here, we have:
\beqn 
f(x,y)&=& n^2 x-\frac{(x+0.5)}{2 r_1^3}-\frac{(x-0.5)}
{2 r_2^3}+\frac{x}{r}f_b(r)=0,\label{eq:gg1jiang}\\
g(x,y)&=& n^2 y-\frac{y}{2 r_1^3}-\frac{y}{2 r_2^3}
+\frac{y}{r}f_b(r)=0,\label{eq:gg2jiang}
\eeqn
where $r_1=\sqrt{(x+0.5)^2+y^2}$ and $r_2=\sqrt{(x-0.5)^2+y^2}$.

For convenience, we define 
\beq
h(y)\equiv n^2-\left[\frac{1}{4}+y^2\right]^{-3/2}
+\frac{f_b(r)}{r}\biggm|_{(0,y)} 
\label{eq:h}
\eeq 
and
\beq
k(x)\equiv  n^2 x-\frac{(x+0.5)}{2|x+0.5|^3}-\frac{(x-0.5)}
{2|x-0.5|^3} + \frac{x}{r}f_b(r)\biggm|_{(x,0)}. \label{eq:k}
\eeq
We have the following properties for the case of equal-mass binary:

{\bf  Property {3.1}}\\ 
 {\it  If $(x_e,y_e)$ is an equilibrium point of System (\ref{eq:3body1}), 
then we have: \\
 either (1) $x_e=0$ and $y_e$ satisfies $h(y)=0$  \\
or (2) $x_e$ satisfies $k(x)=0$ and $y_e = 0$.
}

{\it Proof:} Suppose that $(x_e,y_e)$ is an equilibrium point, thus
it satisfies $f(x,y)=0$ and  $g(x,y)=0$.
From Eq.(\ref{eq:gg2jiang}), we have
$$y_e\left[n^2-\frac{1}{2 r_1^3}-\frac{1}{2 r_2^3}
+\frac{f_b(r)}{r}\right]\biggm|_{(x_e,y_e)}=0.$$
Hence, $y_e=0$ or $[n^2-\frac{1}{2 r_1^3}-\frac{1}{2 r_2^3}
+\frac{f_b(r)}{r}]\biggm|_{(x_e,y_e)}=0$.  
We now discuss these two cases separately.

(I) $[n^2-\frac{1}{2 r_1^3}-\frac{1}{2 r_2^3}
+\frac{f_b(r)}{r}]\biggm|_{(x_e,y_e)}=0$:\\ 
Since $(x_e,y_e)$ is an equilibrium point, that is $f(x_e,y_e)=0$, we have:
\beqn
 0&=&f(x_e,y_e)= x_e\left[n^2-\frac{1}{2 r_1^3}-\frac{1}{2 r_2^3}
+\frac{f_b(r)}{r}\right]-\frac{1}{4 r_1^3}
+\frac{1}{4 r_2^3}\non\\
&=&\frac{1}{4}\left[-\frac{1}{r_1^3}+\frac{1}{r_2^3},\right]. \label{eq:con2} 
\eeqn
Thus,  $r_1=r_2$, i.e.
$(x_e+0.5)^2+y_e^2=(x_e-0.5)^2+y_e^2$. Hence $x_e=0$.
We have  $r_1=r_2=\sqrt{1/4+y_e^2}$ and thus, 
$h(y_e)=n^2-[\frac{1}{4}+y_e^2]^{-3/2}+\frac{f_b(r)}{r}\biggm|_{(0,y_e)}
=[n^2-\frac{1}{2 r_1^3}-\frac{1}{2 r_2^3}+\frac{f_b(r)}{r}]\biggm|_{(0,y_e)}
=0$. 
Therefore,  $x_e=0$ and $y_e$ satisfies 
$h(y)=0$ for the equilibrium point $(x_e,y_e)$.

(II) $y_e=0$:\\
 $f(x_e,y_e)= f(x_e,0)= k(x_e)= 0$ for 
the equilibrium point $(x_e,y_e)$.  Thus, 
$x_e$ satisfies $k(x)=0$ and $y_e = 0$. $\diamondsuit$

{\bf  Property {3.2}} \\
\no {\it (A) If  $y_e$ satisfies  $h(y)=0$, then $(0,y_e)$ is 
an equilibrium point of System (\ref{eq:3body1}).\\
\no (B) If $x_e$ satisfies $k(x)=0$, then 
$(x_e,0)$  is an  equilibrium point of System (\ref{eq:3body1}).}

{\it Proof of (A):}
Suppose that $y_e$ is one of the roots of $h(y)$, i.e. $h(y_e)=0$ and 
we set $x_e=0$.
Because $g(0,y_e) = y_e h(y_e) =0$. 
By Eq.(\ref{eq:gg1jiang}),
$f(0,y_e) = - \frac{1}{4}[\frac{1}{4}+y_e^2]^{-3/2}
+ \frac{1}{4}[\frac{1}{4}+y_e^2]^{-3/2} =0.$ 
Thus, $(0,y_e)$ is the equilibrium point of System (\ref{eq:3body1}). 
$\diamondsuit$

{\it Proof of (B):}
Suppose that $x_e$ is one of the roots of $k(x)$, i.e. $k(x_e)=0$ 
and we set $y_e=0$. Since $y_e=0$, it is trivial that $g(x_e,0) =0$.
Because $k(x_e)=0$, $f(x_e,0)=k(x_e)=0$. 
Thus, $(x_e,0)$ is the equilibrium point
of System (\ref{eq:3body1}). $\diamondsuit$ 

Because of  the above two properties, in stead of searching the roots
for two-variable functions $f(x,y)$ and $g(x,y)$ 
to determine the equilibrium points
on the $x-y$ plane, 
we only need to find the roots of one variable functions $h(y)$ and $k(x)$
to get all the equilibrium points of System (\ref{eq:3body1})
when the mass parameter $\mu=0.5$.
Therefore,
we numerically solve both $h(y)=0$ and $k(x)=0$ 
and find out the number of 
equilibrium points for different given parameters.
The numerical scheme of root finding 
is the Van Wijngaarden-Dekker-Brent Method (Brent 1973). This is 
an excellent algorithm recommended by Press et al. (1992).
We set a high level of  
accuracy that the maximum error is  $10^{-8}$ for the locations of 
equilibrium points on both $x$-axis and $y$-axis. 


Figure 2 shows the results on the $r_i-M_b$ plane.
That is, we numerically search the solution of $k(x)=0$ 
and $h(y)=0$ 
for different $(r_i, M_b)$,
where $0.5 < r_i \leq 1$ and $0.001 \leq M_b \leq 0.35$. 
The grid size is 0.01 for $r_i$ and is 0.02 for $M_b$. 
Those $(r_i, M_b)$ with new equilibrium points are marked by full 
triangle points.

We find that the region with new equilibrium points is itself a triangle.
To have new equilibrium points, the mass of the belt has to be 
larger than 0.15 and $r_i$ is somewhere between 0.7 and 0.8.

It is not surprising that there are more
equilibrium points in our system  
because the potential field of whole system is 
more complicated than the classical restricted three-body problems and thus
there are more possibilities that gravitational forces from 
different components
can balance each other.

\section{The Orbits}

Since we have discovered new equilibrium points
for particular regions of $r_i-M_b$ plane for our system,
it would be interesting to investigate the orbital behavior
around these new equilibrium points.

From Figure 2, we notice that when $M_b=0.3$, $r_i=0.7$, we could have 
new equilibrium points. 
We thus use this as a standard case.
At the $y$-axis, we explicitly 
determine the locations of all equilibrium points, which are at 
$(x_e, y_e) = (0, \pm 0.73), (0,\pm 0.77) ,(0,\pm 1.05)$, in addition to the 
usual $L1$ point at $(0,0)$. (We find that all new equilibrium points
are at $y$-axis because of the choice $\mu_{1}=\mu_{2}=0.5$.)

To understand the properties of these equilibrium points at the $y$-axis, 
we numerically calculate the orbits with initial conditions very close to these
points. There are four initial conditions in this paper, i.e.\\ 
(a) Initial Condition L1: $(x,y,u,v)=(\epsilon,0,0,0)$,\\
(b) Initial Condition E1: $(x,y,u,v)=(\epsilon,0.73, 0,0)$,\\
(c) Initial Condition E2: $(x,y,u,v)=(\epsilon,0.77, 0,0)$,\\
(d) Initial Condition E3: $(x,y,u,v)=(\epsilon,1.05, 0,0)$,\\
where $\epsilon=0.001$.




Figure 3 is the orbits on the $x-y$ plane 
for these four initial conditions with $M_b=0.3$. 
There are four panels, i.e. panel (a) is for Initial Condition L1,
 panel (b) is for Initial Condition E1,
 panel (c) is for Initial Condition E2 and 
 panel (d) is for Initial Condition E3.
To have the idea about the disc's influence, we also integrate 
the orbits with these four initial conditions while the disc mass $M_b=0$
as shown in Figure 4.

For Initial Condition L1, 
Figure 3(a) shows that the orbit fills the Roche lobe, i.e. the 
zero-velocity curve passing the L1 point. The test particle is gravitational 
bound to one of the star because it probably 
has less initial total energy than those
ones in Figure 3(b)-(d). Figure 4(a) shows that the result is similar when 
$M_b=0$. 


For Initial Condition E1,
Figure 3(b) shows that the test particle moves randomly at the beginning
and then steadily move out of the region of central binary.
Figure 4(b) shows that the test particle drifts away from the initial
location regularly.

For Initial Condition E2,
Figure 3(c) is similar to Figure 3(b) but the test particles move more 
chaotically in the central region and then move out slowly at the first stage 
and more quickly finally. Figure 4(c) is very similar to Figure 4(b).

For Initial Condition E3,
Figure 3(d) shows that the test particle is always close to the equilibrium 
point and 
moving on a regular periodical orbit. The equilibrium point could be a 
neutral stable point. Figure 4(d) shows that the test particle is on 
a complicated orbit and drifts away from the central region slowly.

We notice that 
the orbits in Figure 3(a) and Figure 4(a) 
are surrounding one star only.
Most other orbits show that the test particles move spirally outward 
from the central region  except the one in Figure 3(d). 

\section{Lyapunov Exponent}

In addition to analyze the orbits themselves,  
we calculate these orbits' Lyapunov Exponents  
to understand how sensitively  
dependent on the initial conditions. 
In general, 
the larger value of Lyapunov Exponent  means 
more sensitively  
dependent on the initial conditions. 

Our procedure to calculate the Lyapunov Exponent was 
from Wolf et al. (1985).
The maximum Lyapunov Exponent is defined by
$\gamma = {\rm lim}_{t\rightarrow\infty} \chi(t) $, where
$\chi(t)$ is the  Lyapunov Exponent Indicator. 
The system is  
chaotic if $\gamma >0 $ or otherwise regular.
However, it is not possible 
to take $t \rightarrow \infty$ in practice.
We numerically determine the  Lyapunov Exponent Indicator $\chi(t)$
and usually plot $\ln \chi(t) $ as function of $\ln t$
after the evolution is followed up to some time.
If the curve (or the envelope of the curve) on the $\ln t - \ln \chi(t)$ plane 
shows a negative constant slope, the system is
regular, otherwise it is chaotic.

Figure 5(a)-(d) are the results of Lyapunov Exponent for  
Initial Condition L1, E1, E2 and E3 respectively.  
There are two curves in each panel, 
where solid curve is the result of $M_b=0.3$ and 
dotted curve is the result of $M_b=0$.

From these results, we notice that the dotted curves in Figure 5(b)-(c)
and the solid curve in Figure 5(d) show negative
constant slopes. Thus, when $M_b=0$, the orbits of Initial Condition E1 and E2
are likely to be regular. When $M_b=0.3$, only the orbit of 
Initial Condition E3 is likely to be regular. This is probably due to that  
the equilibrium point close to Initial Condition E3 is neutral stable. 
Note that the solid 
curves in Figure 5(b)-(c) are shorter because the test particles were ejected
and the calculations were stopped when
the distances were larger than $10^4$ from the central region.

We can also notice that, in Figure 5(a), the solid curve is always 
higher than the dotted curve. That is, the values of Lyapunov Exponent 
Indicator of the orbit with Initial Condition L1 when $M_b=0.3$ 
are always larger than the one of the orbit with Initial Condition L1 
when $M_b=0$. To summarize, the existence of belt seems to make 
the system to have more chaotic orbits. 

\section{Concluding Remarks}

We have provided the equations for a model which includes the gravitational 
influence from 
a belt around the central binary. We find that, in addition to the usual
Lagrange points, there are  new equilibrium points on the $y$-axis when 
each mass of the components of the binary is equal to $0.5$.
To study the orbits around these new equilibrium points, we 
calculate the orbits and their Lyapunov Exponents with four different 
initial conditions.
It seems that the system could have more chaotic orbits when there is a belt.
 
\section*{Acknowledgements}

We are grateful to the National Center for High-performance Computing
for computer time and facilities. 
This work is supported in part 
by the National Science Council, Taiwan, under Ing-Guey Jiang's 
Grants NSC 94-2112-M-008-010 and also Li-Chin Yeh's
Grants NSC 94-2115-M-134-002.

\clearpage
\appendix
\section{Gravitational Force from the Belt $f_b$}

Let $F(\xi)$ be the elliptic integral of the first kind
and assume
\beq
\xi=\frac{2\sqrt{rr'}}{r+r'}.\label{eq:xi2}
\eeq
By  $\sin^2 A=\frac{1}{2}(1-\cos 2A)$, we have 
\beqn
F(\xi)&=&\int^{\pi/2}_0\frac{1}{\sqrt{1-\xi^2\sin^2\phi'}}d\phi' \non\\ 
&=&(r+r') \int^{\pi/2}_0\frac{1}{\sqrt{r^2+r'^2+2rr'\cos (2\phi')}}d\phi'
\non\\ 
&=&\frac{(r+r')}{2} \int^{\pi}_0\frac{1}{\sqrt{r^2+r'^2+2rr'\cos (\theta)}}
d\theta \non \\
&=&\frac{(r+r')}{2} \int^{\pi}_{0}\frac{1}{\sqrt{r^2+r'^2-2rr'\cos (\phi)}}
d\phi.  \label{eq:f_1}
\eeqn 
Therefore, from Equation (\ref{eq:f_1}), we have
\beqn
& &\int^{2\pi}_{0}\frac{1}{\sqrt{r^2+r'^2-2rr'\cos (\phi)}}d\phi \non \\
&=&2\int^{\pi}_{0}\frac{1}{\sqrt{r^2+r'^2-2rr'\cos (\phi)}}d\phi \non \\  
&=&\frac{4F(\xi)}{r+r'}.\label{eq:f_2}
\eeqn

The gravitational potential is 
\beqn
V(r)&=&-\int_{r'}\!\int^{2\pi}_{0}
\frac{\rho'(r')r'}{\sqrt{r^2+r'^2-2rr'\cos (\phi')}}dr'\!d\phi' \non \\
&=&-4\int_{r'} \frac{F(\xi)\rho'(r')r'}{r+r'}dr', 
\eeqn
where Equation (\ref{eq:f_2}) has been used.
Now we differentiate this potential with respect to $r$ to get
 the gravitational force, so 
\beqn
f_b&=&-\frac{\pa V}{\pa r}\non \\
&=&4\int_{r'} \frac{\frac{dF(\xi)}{d\xi}
\frac{\pa \xi}{\pa r}\rho'(r')r'}{r+r'}dr'\non \\
&+&4\int_{r'} 
F(\xi)\rho'(r')r'\frac{\pa}{\pa r}\left(\frac{1}{r+r'}\right)dr'.
\label{eq:dvdr}
\eeqn
Let $E(\xi)$ be the elliptic integral of the second kind  and 
$\xi'=\sqrt{1-\xi^2}$.
Since $F(\xi)$ is the elliptic integral of the first kind, from 
Byrd \& Friedman (1971),
we have
\beq
\frac{d F(\xi)}{d\xi}=\frac{E-\xi'^2F(\xi)}{\xi\xi'^2}=
\frac{E}{\xi(1-\xi^2)}-\frac{F}{\xi}.\label{eq:dfdxi}
\eeq

By Equation (\ref{eq:xi}) and (\ref{eq:dfdxi}), we calculate
\beqn
\frac{dF(\xi)}{d\xi}\frac{\pa \xi}{\pa r}
&=&\left[
\frac{E}{\xi(1-\xi^2)}-\frac{F}{\xi}\right]\sqrt{\frac{r'}{r}}\frac{r'-r}
{(r+r')^2}\non \\
&=& -\frac{E(r+r')}{2r(r-r')}+\frac{F(r-r')}{2r(r+r')}. \label{eq:dfxi_dxidr}
\eeqn 

We substitute Equation (\ref{eq:dfxi_dxidr}) into 
Equation (\ref{eq:dvdr}), so we have
\beqn
f_b&=&-\frac{\pa V}{\pa r}\non \\
&=& -2\int_{r'}\frac{\rho(r')r'}{r}\left[\frac{1}{r-r'}E(\xi)
+\frac{1}{r+r'}F(\xi)
\right]dr'.\label{eq:f_b}
\eeqn

\clearpage
\begin{figure}
   \epsfysize 6.5 in \epsffile{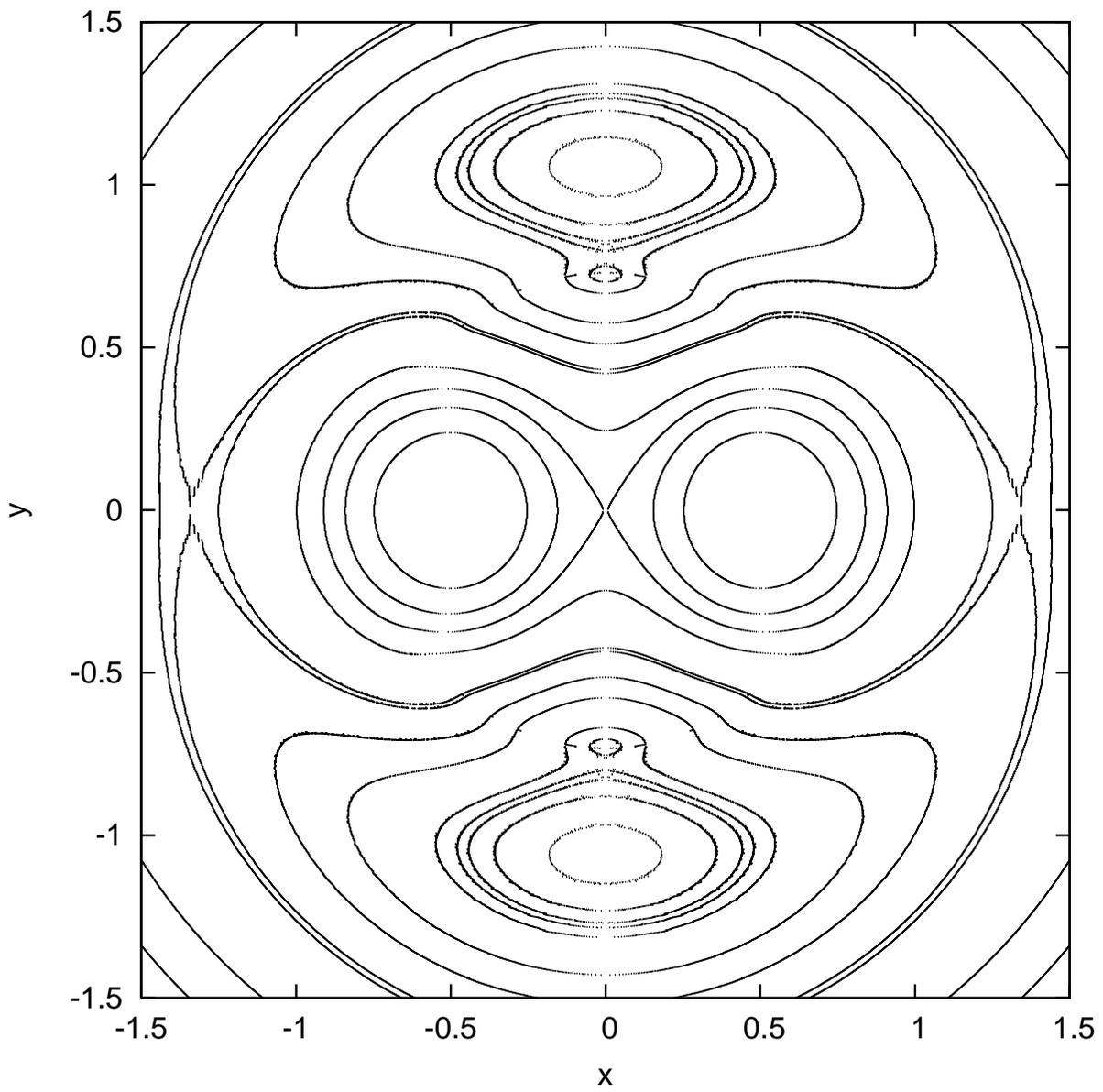}
   \caption{The zero-velocity curves of the system
when $r_i=0.7$ and $M_b=0.3$.
}  
              \label{Fig0}%
    \end{figure}

\clearpage
\begin{figure}
   \epsfysize 6.5 in \epsffile{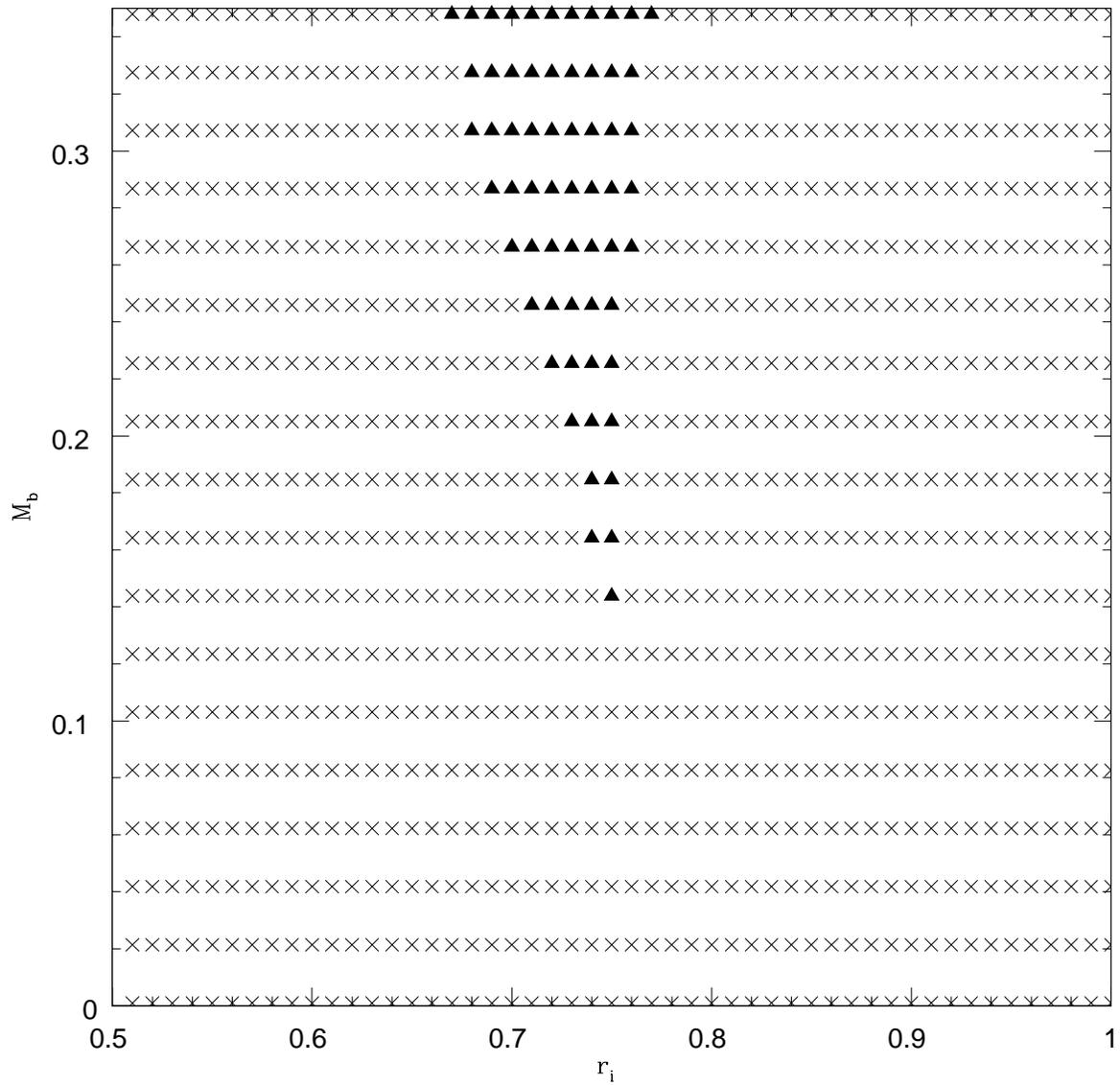}
   \caption{The results of the numerical survey on equilibrium points. 
Those $(r_i, M_b)$ with new equilibrium points are marked by full triangle
points.
}  
              \label{Fig1}%
    \end{figure}

\clearpage
\begin{figure}
\epsfysize 6.5 in \epsffile{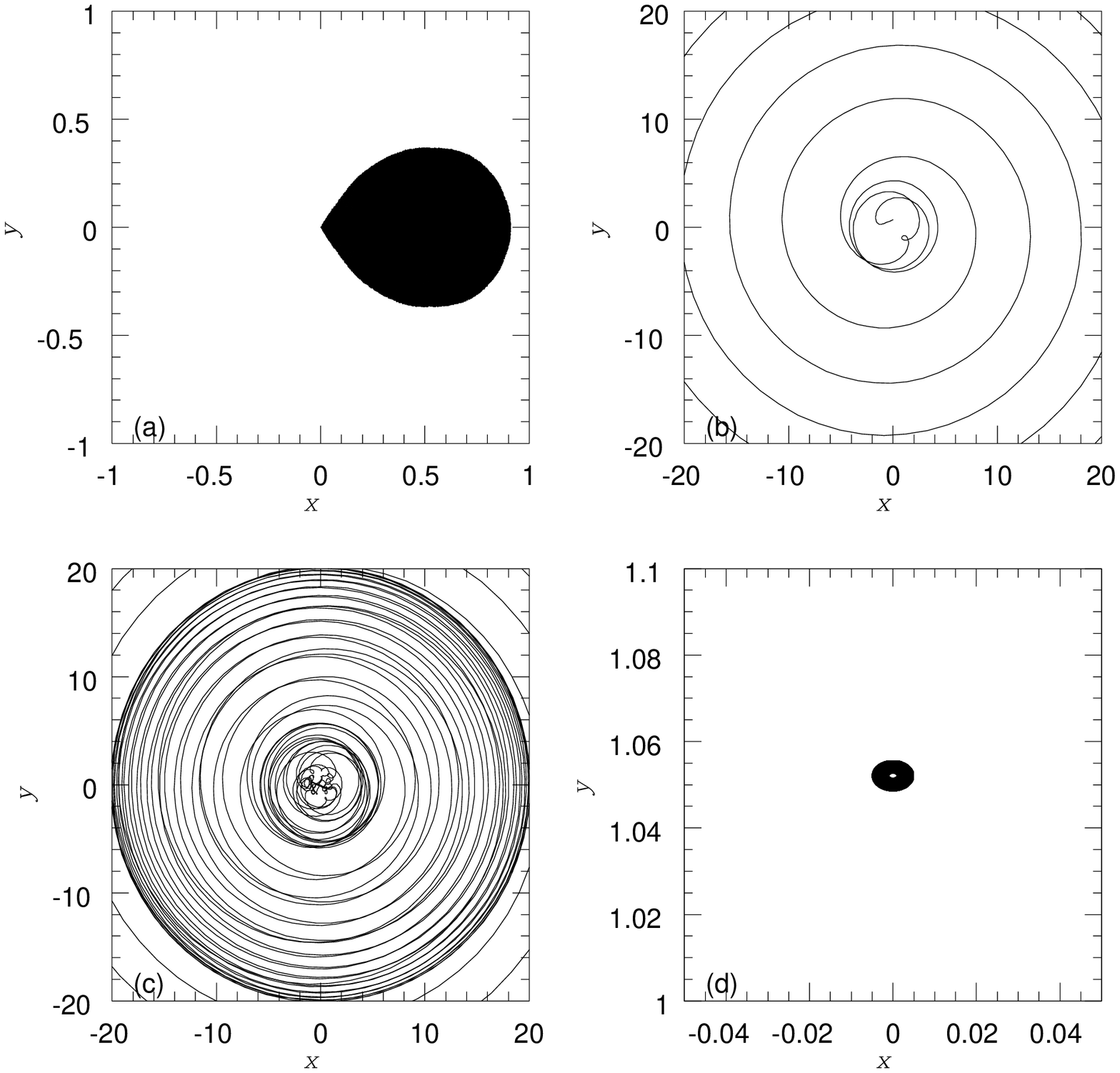}
   \caption{The orbits when the disc mass $M_b=0.3$. Each panel is for
different initial conditions, i.e.
(a) Initial Condition L1;
(b) Initial Condition E1;
(c) Initial Condition E2;
(d) Initial Condition E3.
}  
\label{Fig2}
\end{figure}
\clearpage
\begin{figure}
\epsfysize 6.5 in \epsffile{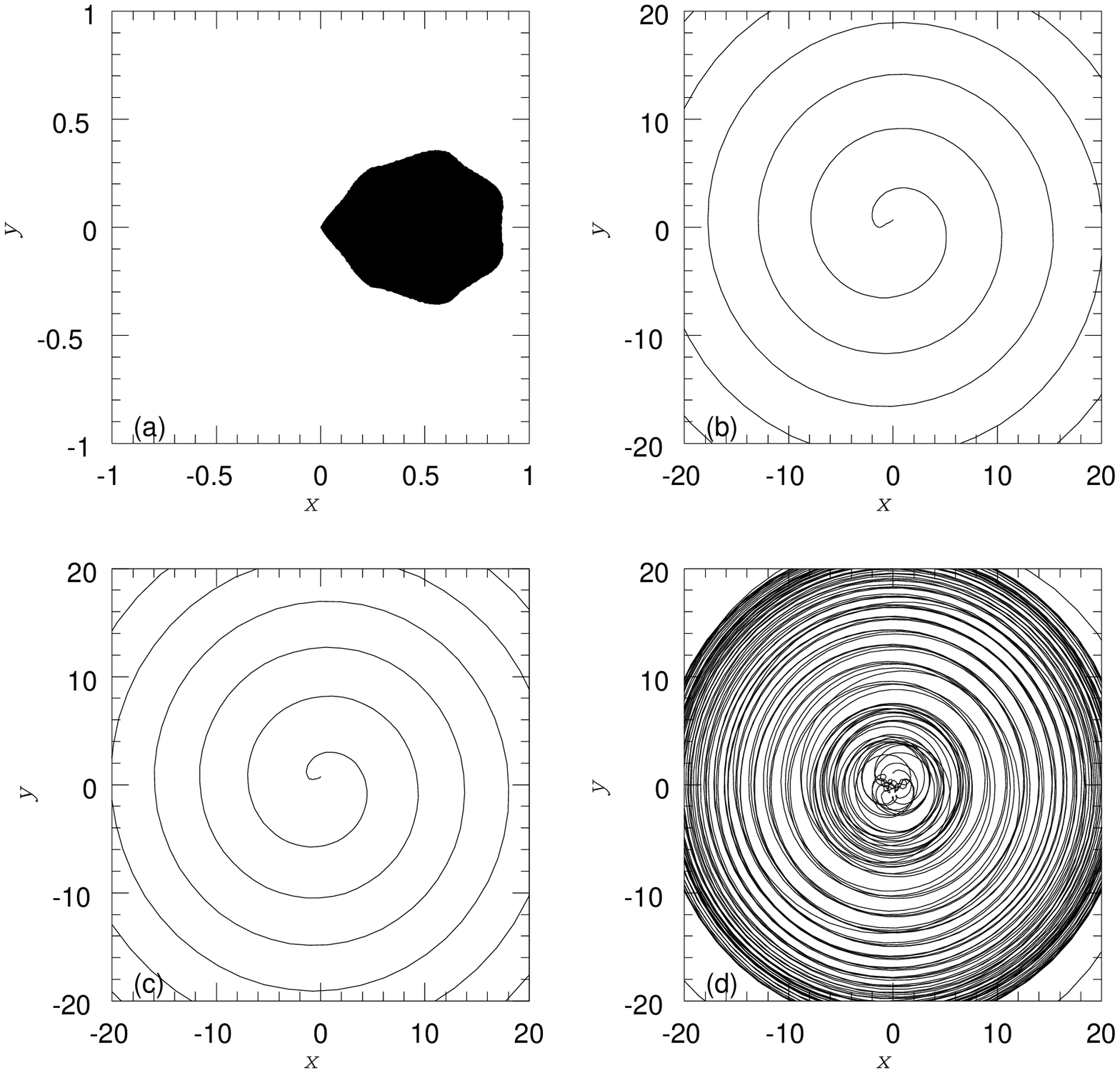}
   \caption{The orbits when the disc mass $M_b=0$. Each panel is for
different initial conditions, i.e.
(a) Initial Condition L1;
(b) Initial Condition E1;
(c) Initial Condition E2;
(d) Initial Condition E3.
}  
\label{Fig3}
\end{figure}

\clearpage
 \begin{figure}
\epsfysize 6.5 in \epsffile{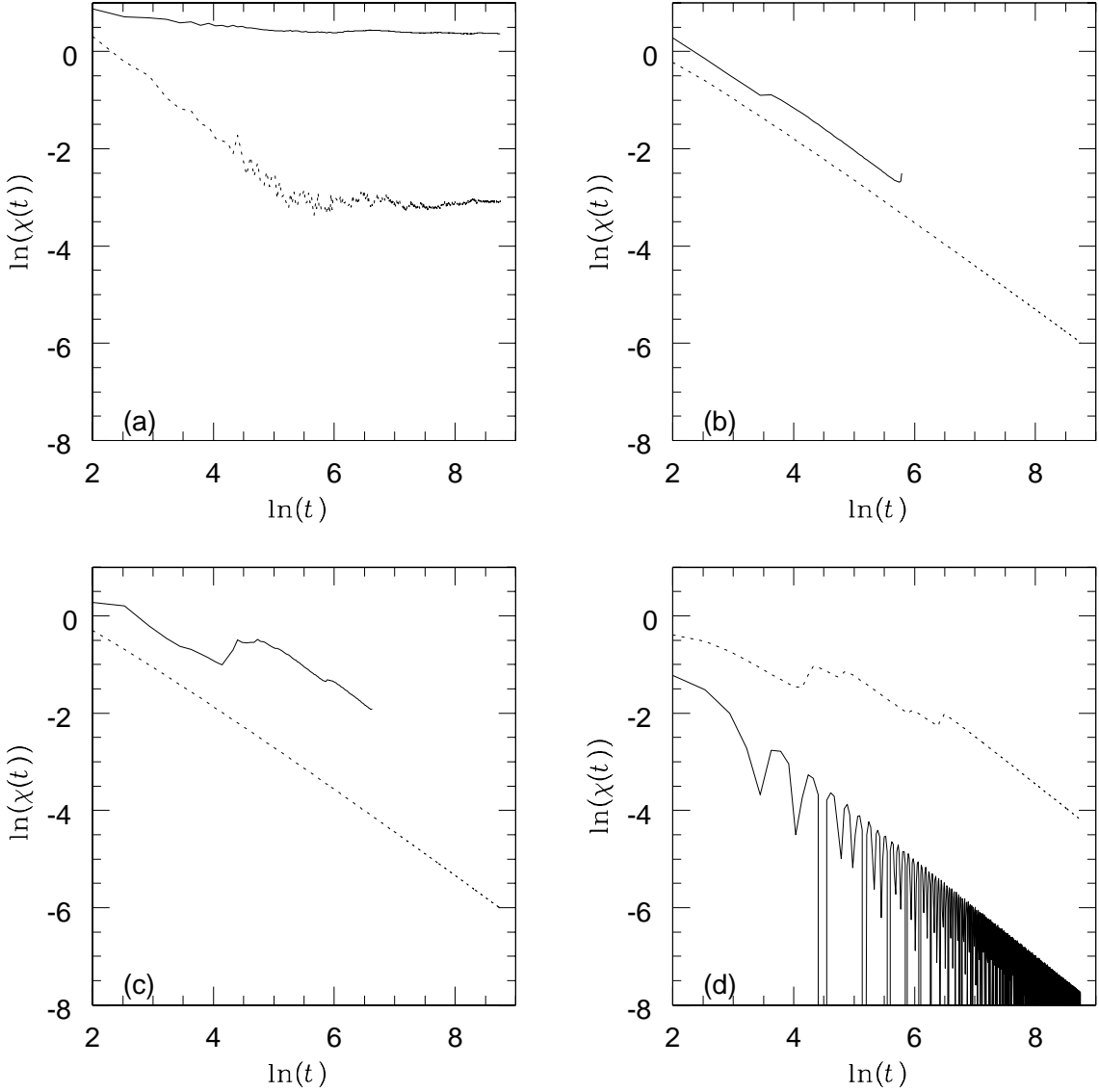}
   \caption{Lyapunov Exponent 
for different disc mass and initial condition, i.e. 
Panel (a) is for Initial Condition L1;
Panel (b) is for Initial Condition E1;
Panel (c) is for Initial Condition E2;
Panel (d) is for Initial Condition E3.
In each panel, solid curve is for $M_b=0.3$, dotted curve is for 
$M_b=0$.
}  
\label{Fig4}
\end{figure}

\end{document}